\documentclass[aps,pra,showpacs]{revtex4}

\usepackage{graphicx}
\usepackage{amsmath}
\usepackage{amssymb}

\newcommand{\al}{\alpha}
\newcommand{\la}{\lambda}
\newcommand{\prt}{\partial}
\newcommand{\om}{\omega}

\begin{document}

\title{Expansion of Bose-Einstein condensates in lower dimensions}

\author{A.M. Kamchatnov}
\email{kamch@isan.troitsk.ru}

\affiliation{
Institute of Spectroscopy, Russian Academy of Sciences, Troitsk 142190,
Moscow Region, Russia\\
and Instituto de F\'{\i}sica Te\'{o}rica, UNESP,
Rua Pamplona 145, 01405-900 S\~{a}o Paulo, Brazil
}

\date{\today}


\begin{abstract}
In the hydrodynamic approximation we obtain analytic solutions to the
Gross-Pitaevskii equation with positive scattering length which describe
expansions of the Bose-Einstein condensates in quasi-one and quasi-two
dimensional geometries. The expansion laws are expressed in terms of
the initial sizes of the condensate and the trap frequencies before
release, that is in terms of experimentally measurable parameters
only. Three-dimensional effects are estimated with the use of
variational approach.  The analytical formulae show good agreement
with available experimental data.
\end{abstract}

\maketitle

\section{Introduction}

The properties of Bose-Einstein condensates (BECs) in lower-dimensional
geometries, when the particles motion is tightly confined to zero point
oscillations in one or two directions, have attracted much
attention \cite{Bongs01}-\cite{PS03}
because of specific coherence properties of
degenerate Bose gases in lower dimensions and possible experimental
realization of quantum gases beyond mean field approach. In order to realize
a one-dimensional (1D) condensate, optical dipole traps have been used to
achieve necessary asymmetry between the trapping frequencies (see, e.g.
\cite{Bongs01}).
Two-dimensional (2D) condensates have been created in an array of
pancake-shaped traps formed by the periodic potential of a 1D optical
lattice \cite{Morsch02}.
For large lattice potential depths, the motion along the
direction of the array is frozen and the condensate is effectively
split up into several smaller independent condensates confined in the
individual lattice wells.

Much experimental information on BECs in three-dimensional confining
traps has been provided by imaging the density of the atomic cloud after
switching off the trap potential. In the mean-field approximation,
evolution of BECs in diluted
gases is described by the time-dependent Gross-Pitaevskii (GP) equation
\cite{PS03}
\begin{equation}\label{GP}
i\hbar\frac{\prt\psi}{\prt t}=-\frac{\hbar^2}{2m}\Delta\psi+
V_{ext}(\mathbf{r})\psi+g|\psi|^2\psi,
\end{equation}
where
\begin{equation}\label{pot}
V_{ext}(\mathbf{r})=\frac12m(\om_x^2x^2+\om_y^2y^2+\om_z^2z^2)
\end{equation}
is the external trap potential,
$g={4\pi\hbar^2a_s}/m$
is the nonlinear interaction constant
$a_s$ being the $s$-wave scattering length of the interacting atoms, and
the condensate wave function $\psi$ is normalized to the number of atoms
\begin{equation}\label{norm3}
\int|\psi|^2d\mathbf{r}=N.
\end{equation}
If the number of atoms is large enough, the GP equation can be
reduced to the hydrodynamic equations, which admit simple
self-similar solutions describing as oscillations of BEC in
the parabolic trap potential, so its free 3D expansion
after switching off the trap potential \cite{CastinDum96}-\cite{Dalfovo97}.
This theory shows excellent agreement with the experiment.

In the recent experiments the free expansion of BEC has been
investigated in 1D waveguides \cite{Bongs01} and in the array of 2D
pancake BECs \cite{Morsch02}, and this line of investigations looks
very promising.
In \cite{SZ02} expansion of BEC in a
waveguide has been considered for quasi-1D case but without
taking into account the quantum pressure in transverse direction.
In \cite{Plaja02} the effects of the quantum pressure have been
calculated for stationary states with account of two transverse
modes and expansion dynamic has been studied numerically.
In \cite{Das02-2} stationary states of BEC in cigar-shaped and pancake
traps have been considered by means of proper variational
approach without discussion of free expansion dynamics.
The aim of this article is to give
analytic treatment of the lower dimensional expansion of BEC.
To this end, under conditions that 3D GP equation
can be reduced to its lower dimensional counterparts, in Section II
we solve
these reduced equations in their hydrodynamic approximations
with initial conditions corresponding to the equilibrium
states of a confined in the parabolic trap BEC with positive scattering
length ($a_s>0$). Free expansion of BEC takes
place either in longitudinal direction of the 1D waveguide,
or in transverse direction of the 2D pancake trap.
 If the conditions of lower
dimensional reductions of GP equation are not fulfilled, then
effects of 3D hydrodynamic flow can be estimated with the use
of variational approach what is done in
Section III. Section IV is devoted to comparison of the theory
with experiment.

\section{Free 1D and 2D expansion of BEC}

The GP equations admits formulation in the form of the Hamilton
least action principle with the action functional
\begin{equation}\label{action}
S=\int Ldt, \qquad L=\int \mathcal{L}d\mathbf{r},
\end{equation}
where the Lagrangian density is given by
\begin{equation}\label{3Lagr-density}
\mathcal{L}=\frac{i\hbar}2(\psi_t^*\psi-\psi_t\psi^*)+
\frac{\hbar^2}{2m}|\nabla\psi|^2+V_{ext}|\psi|^2+\frac12g|\psi|^4,
\end{equation}
so that the corresponding Lagrange equation
coincides with the GP equation (\ref{GP}). In cigar-shaped or pancake
configurations it is not difficult to determine conditions when
the tightly confined degrees of freedom become irrelevant
(``frozen'') and the GP equation reduces to the effective 1D or
2D equation. This problem has been
considered in many papers (see, e.g. \cite{PS03}, Chapter 17, and
references therein) and we shall present
here only the main points with the aim to define the relevant parameters
and to formulate the applicability conditions.

\subsection{One-dimensional expansion}

If the axial frequency $\om_z$ of the axially symmetric trap is
much less than its transverse frequency $\om_\bot$,
\begin{equation}\label{la1}
\la=\frac{\om_z}{\om_\bot}\ll 1,
\end{equation}
and characteristic energy of zero-point transverse
motion is much greater than the nonlinear energy, then the
transverse motion reduces to the ground state of quantum oscillations
in the radial trap potential. Characteristic radius of this motion
is equal to $a_\bot=(\hbar/m\om_\bot)^{1/2}$, and if we denote by
$Z_0$ the characteristic axial size of the BEC, then with account of
the estimate $N\sim|\psi|^2a_\bot^2Z_0$ (see Eq.~(\ref{norm3})) the
corresponding condition
can be written in the form (see, e.g. \cite{Menotti02})
\begin{equation}\label{1D-cond1}
\frac{Na_s}{Z_0}\ll1.
\end{equation}
If this condition is fulfilled, then we can factorize the condensate
wave function
\begin{equation}\label{1D-factor}
\psi(\mathbf{r},t)=\phi(x,y)\Psi(z,t),
\end{equation}
where
\begin{equation}\label{1D-osc}
\phi(x,y)=\frac1{\sqrt{\pi}a_\bot}\exp\left(-\frac{x^2+y^2}
{2a_\bot^2}\right),
\end{equation}
is the ground state wave function of the transverse motion.
Substitution of (\ref{1D-factor}),(\ref{1D-osc}) into
(\ref{action}),(\ref{3Lagr-density}) and subsequent integration
over transverse
cross section lead to the expression for the action in terms of
effective 1D Lagrangian density
\begin{equation}\label{1Lagr-density}
\mathcal{L}_{1D}=\frac{i\hbar}2(\Psi_t^*\Psi-\Psi_t\Psi^*)+
\frac{\hbar^2}{2m}|\Psi_z|^2
+\frac12m\om_z^2z^2|\Psi|^2+\frac{g}
{4\pi a_\bot^2}|\Psi|^4
\end{equation}
(irrelevant constant term has been omitted). Now the evolution of
$\Psi(z,t)$ is governed by Lagrange equation
which yields the 1D GP equation
\begin{equation}\label{1D-GP}
i\hbar\Psi_{t}=-\frac{\hbar^2}{2m}\Psi_{zz}+
\frac12m\om_z^2z^2\Psi+g_{1D}|\Psi|^2\Psi,
\end{equation}
where
\begin{equation}\label{g1D}
g_{1D}=\frac{g}{2\pi a_\bot^2}=\frac{2\hbar^2 a_s}{ma_\bot^2}
\end{equation}
is effective 1D nonlinear coupling constant
and $\Psi$ is normalized according to
\begin{equation}\label{norm1}
\int|\Psi|^2dz=N.
\end{equation}
Equation (\ref{1D-GP}) governs 1D longitudinal dynamics of
the BEC in cigar-shaped trap.

By means of well-known substitution
\begin{equation}\label{Madelung}
\Psi(z,t)=\sqrt{\rho(z,t)}\exp\left(\frac{im}{\hbar}\int^z
v(z',t)dz'\right)
\end{equation}
equation (\ref{1D-GP}) transforms to the system
\begin{equation}\label{1D-cont}
\rho_t+(\rho v)_z=0,
\end{equation}
\begin{equation}\label{1D-quantum}
v_t+vv_z+\frac{g_{1D}}m\rho_z+\om_z^2z+\frac{\hbar^2}{2m^2}
\left(\frac{\rho_z^2}{4\rho^2}-\frac{\rho_{zz}}{2\rho}\right)_z=0.
\end{equation}
The last term in Eq.~(\ref{1D-quantum})
called ``quantum pressure'' can be neglected, if
it is much less than the nonlinear one proportional to $g_{1D}$,
that is the condition
\begin{equation}\label{1D-cond2}
\frac{a_\bot}{Z_0}\ll\frac{Na_s}{a_\bot}
\end{equation}
is fulfilled. Then Eq.~(\ref{1D-quantum}) reduces to
\begin{equation}\label{1D-Euler}
v_t+vv_z+\frac{g_{1D}}m\rho_z+\om_z^2z=0
\end{equation}
which together with Eq.~(\ref{1D-cont}) provides the hydrodynamic
approximation to description of BEC dynamics in parabolic trap
potential.

The stationary solution of the hydrodynamic equations gives the
well-known Thomas-Fermi (TF) distribution of density
for 1D BEC:
\begin{equation}\label{1D-TF}
\rho(z)=\frac{3N}{4Z_0}\left(1-\frac{z^2}{Z_0^2}\right),
\quad v=0,
\end{equation}
where the integration constant $Z_0$ has a meaning of the axial half-length
of the condensate and can be expressed in terms of the number of
atoms $N$,
\begin{equation}\label{Z}
Z_0=(3Na_sa_\bot^2\la^{-2})^{1/3}.
\end{equation}
Conditions (\ref{1D-cond1}) and (\ref{1D-cond2}) of applicability of
1D hydrodynamic approximation can be written upon substitution of
(\ref{Z}) as (see \cite{Menotti02})
\begin{equation}\label{1D-cond}
\sqrt{\la}\ll\frac{Na_s}{a_\bot}\ll\frac1\la.
\end{equation}

Now we suppose that the axial trap is turned off, so that BEC expands
freely in the axial direction. At the same time it remains being
confined in the transverse direction, that is the radial motion is still
``frozen'' to the quantum ground state (\ref{1D-osc}). Hence, this
expansion is governed by the hydrodynamic equations (\ref{1D-cont})
and (\ref{1D-Euler}) with the initial conditions (\ref{1D-TF}). Similar
problems were studied in nonlinear optics \cite{Talanov} for the opposite
sign of the pressure term $\rho_z$ in (\ref{1D-Euler})
and for description of
free 3D expansion of BEC \cite{CastinDum96,Kagan96-1,Kagan96-2,Dalfovo97}.
Here we use
the same approach to the 1D expansion of BEC in a ``waveguide''. We
look for the solution of Eqs.~(\ref{1D-cont}),(\ref{1D-Euler}) in
the form
\begin{equation}\label{1D-ansatz}
\rho(z,t)=\frac{3N}{4Z_0}\frac1{b_z(t)}
\left(1-\frac{z^2}{Z_0^2b_z^2(t)}\right),
\quad v(z,t)=z\al_z(t),
\end{equation}
where $b_z(t)$ and $\al_z(t)$ must satisfy the initial conditions
\begin{equation}\label{1D-init}
b_z(0)=1, \qquad \al_z(0)=0.
\end{equation}
Substitution of (\ref{1D-ansatz}) into (\ref{1D-cont}),(\ref{1D-Euler})
yields the relation
\begin{equation}\label{al-z}
\al_z(t)=\dot{b}_z(t)/b_z(t)
\end{equation}
and the equation for $b_z(t)$
\begin{equation}\label{b-zdyn}
\ddot{b}_z=\om_z^2/b_z^2
\end{equation}
which can be integrated at once to give (see \cite{BKK}, where this
solution was applied to the description of the initial quasi-1D stage
of expansion of pancake BEC in axial direction)
\begin{equation}\label{b-z}
\sqrt{2}\om_zt=\sqrt{b_z(b_z-1)}+\frac12\ln\left[2b_z-1+
2\sqrt{b_z(b_z-1)}\right].
\end{equation}
This formula determines implicitly $b_z$ as a function of time $t$.
The function $\al(t)$ can be easily expressed in terms of $b_z(t)$,
so that velocity field is given by
\begin{equation}\label{1D-v}
v(z,t)=\frac{\sqrt{2}\om_zz}{b_z(t)}\sqrt{1-\frac1{b_z(t)}}.
\end{equation}
The edge point of the density distribution moves according to the law
\begin{equation}\label{zmax}
z_{max}(t)=Z_0b_z(t)
\end{equation}
with the maximal velocity
\begin{equation}\label{vmax}
v_{max}(t)=\frac{d{z}_{max}}{dt}=\sqrt{2}Z_0\om_z\sqrt{1-\frac1{b_z(t)}}.
\end{equation}
In the limit of large $t\gg\om_z^{-1}$ we find
\begin{equation}\label{as-bz}
b_z(t)\simeq\sqrt{2}\,\om_zt, \qquad t\gg\om_z^{-1},
\end{equation}
so that distributions of density and velocity fields get simple form
\begin{equation}\label{1D-asymp}
\rho(z,t)\simeq\frac{3N}{4v_{max}}\frac1{t}
\left(1-\frac{z^2}{v_{max}^2t^2}\right),
\quad v(z,t)\simeq \frac{z}t, \quad t\gg\om_z^{-1},
\end{equation}
and maximal velocity becomes constant
\begin{equation}\label{vmax-as}
v_{max}\simeq \sqrt{2}Z_0\om_z,\qquad t\gg\om_z^{-1}.
\end{equation}
These formulae describe a hydrodynamic flow ``by inertia'' when
the density becomes so small that the nonlinear pressure does not
accelerate the gas anymore. The formula (\ref{vmax-as}) is
convenient for comparison with experiment since the asymptotic
value of the maximal velocity is expressed in terms of two
experimentally measurable parameters---the axial frequency $\om_z$ of
the trap initially confining the BEC and the initial half-width $Z_0$
of the axial TF distribution.

From (\ref{1D-asymp}) we find the asymptotic distribution function
of velocities
\begin{equation}\label{1D-vel-asymp}
\rho(v)dv=\frac{3N}{4v_{max}}\left(1-\frac{v^2}{v_{max}^2}\right)dv,
\quad |v|\leq v_{max}.
\end{equation}
The mean kinetic energy (``release'' energy) is given by
\begin{equation}\label{1D-release}
\overline{E}=\frac{m}{2N}\int v^2\rho(v)dv=\frac1{5}E_{max},
\quad E_{max}=\frac12mv_{max}^2.
\end{equation}

\subsection{Two-dimensional expansion}

We can achieve 2D dynamics of BEC, if the axial frequency $\om_z$
of the trap is much greater than its radial frequency $\om_\bot$,
that is the inequality (\ref{la1}) is replaced by the inverse one,
\begin{equation}\label{la2}
\la=\frac{\om_z}{\om_\bot}\gg 1.
\end{equation}
In this case we assume that the axial motion along the $z$ axis
is ``frozen'', that is the energy of its zero point oscillatory
motion with the quantum amplitude $a_z=(\hbar/m\om_z)^{1/2}$ is
much greater than the nonlinear energy. This condition with the
use of an estimate $N\sim|\psi|^2R_0^2a_z$, where $R_0$ denotes a
characteristic radius of the density distribution in the pancake
$(x,y)$ plane, leads to the inequality
\begin{equation}\label{2D-cond1}
\frac{Na_s}{a_z}\ll\left(\frac{R_0}{a_z}\right)^2.
\end{equation}
If this condition is fulfilled, then we can again factorize the
condensate wave function:
\begin{equation}\label{2D-factor}
\psi(\mathbf{r},t)=\phi(z)\Psi(x,y,t),
\end{equation}
where
\begin{equation}\label{2D-osc}
\phi(z)=\frac1{\pi^{1/4}a_z^{1/2}}\exp\left(-\frac{z^2}
{2a_z^2}\right)
\end{equation}
is the ground state wave function of the axial motion.
Substitution of (\ref{2D-factor}),(\ref{2D-osc}) into (\ref{action}),
(\ref{3Lagr-density}) and integration over longitudinal coordinate
lead to the expression for the action in terms of the
effective 2D Lagrangian density
\begin{equation}\label{2Lagr-density}
\mathcal{L}_{2D}=\frac{i\hbar}2(\Psi_t^*\Psi-\Psi_t\Psi^*)+
\frac{\hbar^2}{2m}(|\Psi_x|^2+|\Psi_y|^2)
+\frac12m\om_\bot^2(x^2
+y^2)|\Psi|^2+\frac{g}
{2\sqrt{2\pi} a_z}|\Psi|^4.
\end{equation}
The corresponding Lagrange equation yields the 2D GP equation
\begin{equation}\label{2D-GP}
i\hbar\Psi_{t}=-\frac{\hbar^2}{2m}\Delta_{\bot}\Psi+
\frac12m\om_\bot^2(x^2+y^2)\Psi+g_{2D}|\Psi|^2\Psi,
\end{equation}
where $\Delta_\bot=\prt_x^2+\prt_y^2$ is the transverse Laplace
operator, $g_{2D}$ is the effective nonlinear interaction constant,
\begin{equation}\label{g2D}
g_{2D}=\frac{g}{\sqrt{2\pi} a_z}=\frac{2\sqrt{2\pi}\hbar^2 a_s}{ma_z},
\end{equation}
and $\Psi$ is normalized according to
\begin{equation}\label{norm2}
\int|\Psi|^2dxdy=N.
\end{equation}
The NLS equation (\ref{2D-GP}) governs 2D transverse dynamics of
the BEC in a pancake trap.

By means of a substitution
\begin{equation}\label{Madelung2}
\Psi(\mathbf{r}_\bot,t)=\sqrt{\rho(\mathbf{r}_\bot,t)}\exp\left(
\frac{im}\hbar\int^{\mathbf{r}_\bot}\mathbf{v}(\mathbf{r}_\bot',t)
\cdot d\mathbf{r}'\right),
\end{equation}
where $\mathbf{r}_\bot=(x,y)$ and velocity $\mathbf{v}=(v_x,v_y)$
has components only in the
transverse plane, the equation (\ref{2D-GP}) transforms to the system
\begin{equation}\label{2D-cont}
\rho_t+\nabla_\bot(\rho\mathbf{v})=0,
\end{equation}
\begin{equation}\label{2D-quantum}
\mathbf{v}_t+(\mathbf{v}\nabla_\bot)\mathbf{v}+\frac{g_{2D}}m
\nabla_\bot\rho+\om_\bot^2\mathbf{r}
+\frac{\hbar^2}{2m^2}
\nabla_\bot\left[\frac{(\nabla_\bot\rho)^2}{4\rho^2}-
\frac{\Delta_\bot\rho}{2\rho}\right]=0,
\end{equation}
where $\nabla_\bot=(\prt_x,\prt_y)$ is the transverse gradient operator.

The quantum pressure term can be neglected if it is much less than the
nonlinear term what leads to the condition
\begin{equation}\label{2D-cond2}
1\ll\frac{Na_s}{a_z}.
\end{equation}
Then Eq.~(\ref{2D-quantum}) takes the form of the hydrodynamic equation
\begin{equation}\label{2D-Euler}
\mathbf{v}_t+(\mathbf{v}\nabla_\bot)\mathbf{v}+\frac{g_{2D}}m
\nabla_\bot\rho+\om_\bot^2\mathbf{r}=0.
\end{equation}
The stationary solution of hydrodynamic equations corresponds to 2D TF
approximation
\begin{equation}\label{2D-TF}
\rho(r)=\frac{2N}{\pi R_0^2}\left(1-\frac{r^2}{R_0^2}\right),
\quad \mathbf{v}=0,
\end{equation}
where $r^2=x^2+y^2$ and the radius $R_0$ of the density distribution
is determined by the number of atoms $N$:
\begin{equation}\label{R}
R_0=\left(\frac{16}{\sqrt{2\pi}}Na_sa_z^3\la^2\right)^{1/4}.
\end{equation}
It permits one to transform the inequality (\ref{2D-cond1}) to more
convenient form, so that the applicability conditions of 2D
approximation in the TF limit become
\begin{equation}\label{2D-cond}
1\ll \frac{Na_s}{a_z}\ll \la^2.
\end{equation}

When the radial trap is turned off, the BEC expands in the radial
direction, but remains confined in the axial one. This radial
expansion is described by the hydrodynamic equations (\ref{2D-cont})
and (\ref{2D-Euler}) with the initial condition (\ref{2D-TF}). Now
we look for the solution in the form
\begin{equation}\label{2D-ansatz}
\rho(r,t)=\frac{2N}{\pi R_0^2}\frac1{b_\bot^2(t)}
\left(1-\frac{r^2}{R_0^2b_\bot^2(t)}\right),
\quad {v}(r,t)=r\al_\bot(t),
\end{equation}
where $v$ is the radial component of velocity, and $b_\bot(t)$
and $\al_\bot(t)$ must satisfy the initial conditions
\begin{equation}
b_\bot(0)=1,\qquad \al_\bot(0)=0.
\end{equation}
Substitution of (\ref{2D-ansatz}) into (\ref{2D-cont}),(\ref{2D-Euler})
leads to the relation $\al_\bot=\dot{b}_\bot/b_\bot$ between $b_\bot(t)$
and $\al_\bot(t)$ and to the differential equation for $b_\bot(t)$:
\begin{equation}\label{2D-eq}
\ddot{b}_\bot=\om_\bot^2/b_\bot^3.
\end{equation}
Its integration with the initial conditions $b_\bot(0)=1$,
$\dot{b}_\bot(0)=b_\bot(0)\al_\bot(0)=0$ gives
\begin{equation}\label{2D-b}
b_\bot(t)=\sqrt{1+\om_\bot^2t^2},
\end{equation}
and hence $\al_\bot(t)=\dot{b}_\bot/b_\bot=\om_\bot^2t/(1+\om_\bot^2t^2)$.
Thus we obtain simple formulae for radial distributions of density and
velocity fields:
\begin{equation}\label{2D-sol}
\rho(r,t)=\frac{2N}{\pi R_0^2}\frac1{1+\om_\bot^2t^2}
\left(1-\frac{r^2}{R_0^2(1+\om_\bot^2t^2)}\right),\quad
 {v}(r,t)=\frac{\om_\bot^2rt}{1+\om_\bot^2t^2}.
\end{equation}
The edge point of the radial density distribution moves according to the law
\begin{equation}\label{rmax}
r_{max}(t)=R_0\sqrt{1+\om_\bot^2t^2}
\end{equation}
with maximal velocity
\begin{equation}\label{2Dvmax}
v_{max}(t)=\frac{dr_{max}}{dt}=\frac{R_0\om_\bot^2t}{\sqrt{1+\om_\bot^2t^2}}.
\end{equation}

In the limit of asymptotically large time $t\gg \om_\bot^{-1}$ we find
\begin{equation}\label{2D-asymp}
\rho(r,t)\simeq\frac{2N}{\pi v_{max}}\frac1{t^2}
\left(1-\frac{r^2}{v_{max}^2t^2}\right),
\quad {v}(r,t)\simeq\frac{r}{t},
\end{equation}
where
\begin{equation}\label{asymp}
v_{max}\simeq R_0\om_\bot, \qquad t\gg \om_\bot^{-1}.
\end{equation}
As analogous formulae in 1D case, these formulae describe a flow
``by inertia''.
Again the maximal velocity is expressed in
terms of two experimentally measurable parameters---the initial
radial frequency $\om_\bot$ of the trap before its switching off
and the initial radius $R_0$ of the TF distribution.

The asymptotic distribution of velocities is given by
\begin{equation}\label{a}
\rho(v)dv=\frac{4N}{v_{max}^2}\left(1-\frac{v^2}{v_{max}^2}\right)vdv,
\end{equation}
and the release energy is equal to
\begin{equation}\label{b}
\overline{E}=\frac13E_{max},\quad E_{max}=\frac12mv_{max}^2.
\end{equation}

\section{Effects of 3D flow: Variational approach}

When conditions (\ref{1D-cond1}) or (\ref{2D-cond1}) are not fulfilled,
the motion along the shortest size of the condensate cannot be
considered as frozen one and we have to include into consideration
the radial (for cigar-shaped trap) or axial (for pancake trap) flow.
As was shown in \cite{Perez96}, it can be done by means of simple
variational approach. However, in our case of strongly asymmetrical
traps it is more natural for large enough number of atoms to assume
the TF rather than Gaussian distribution along longer direction of the
condensate, as it was done for static equilibrium case in
\cite{Das02-2}. Here we apply similar approach to dynamics of BEC.

\subsection{Cigar-shaped trap}

For cigar-shaped trap with TF distribution of density along the axial
$z$ direction we take the variational trial function in the form
\begin{equation}\label{1D-trial}
\psi=A\exp\left(-\frac{r^2}{2w_\bot^2}\right)\sqrt{1-\frac{z^2}
{w_z^2}}\exp\left[\frac{i}2(\al_\bot^2r^2+\al_z^2z^2)\right],
\end{equation}
where $A,\,w_\bot,\,w_z,\,\al_\bot,\,\al_z$ are assumed to be
functions of time. It is supposed that $Na_s/Z_0\sim 1$, that is
in the radial direction the TF distribution is not achieved yet
and therefore the radial wavefunction can be approximated well
enough by the Gaussian function. The variable $A$ is connected
with widths $w_\bot$, $w_z$ by the normalization condition
(\ref{norm3}) which yields
\begin{equation}\label{1D-A}
A=\left(\frac{3N}{4\pi w_\bot^2w_z}\right)^{1/2}.
\end{equation}
Substitution of (\ref{1D-trial}) into (\ref{action}),(\ref{3Lagr-density})
yields after integration the averaged Lagrangian
\begin{equation}\label{1D-Lagr-av}
\begin{split}
\frac{L}{N}=\left[\frac{\hbar^2}2\frac{d\al_\bot}{dt}+\frac{\hbar^2}{2m}
\left(\frac1{w_\bot^4}+\al_\bot^2\right)+\frac12m\om_\bot^2\right]w_\bot^2\\
+ \left(\frac{\hbar^2}2\frac{d\al_z}{dt}+\frac{\hbar^2}{2m}
\al_z^2+\frac12m\om_\bot^2\la^2\right)\frac{w_z^2}5+
\frac{3Na_s\hbar^2}{5m}\frac1{w_\bot^2w_z},
\end{split}
\end{equation}
where we have neglected the term $1/w_z^4$ compared with $\al_z^2$
since according to the estimates $w_z\sim Z_0$, $\al_z\sim
(m/\hbar)\dot{w}_z/w_z\sim m\om_z/\hbar\sim\la/a_\bot^2$ the
condition $1/w_z^2\ll\al_z$ coincides exactly with the condition
$\la^{1/2}\ll Na_s/a_\bot$ of applicability of the TF approximation
in the axial direction (see (\ref{1D-cond})).

The Lagrangian (\ref{1D-Lagr-av}) leads to well-known formulas
\begin{equation}\label{al}
\al_\bot=\frac{m}{\hbar}\frac1{w_\bot}\frac{dw_\bot}{dt},\quad
\al_z=\frac{m}{\hbar}\frac1{w_z}\frac{dw_z}{dt},
\end{equation}
and to equations of motion for widths:
\begin{equation}\label{wbot-eq}
\ddot{w}_\bot+\om_\bot^2w_\bot=\frac{\hbar^2}{m^2w_\bot^3}+
\frac{6Na_s\hbar^2}{5m^2}\frac1{w_\bot^3w_z}
\end{equation}
\begin{equation}\label{wz-eq}
\ddot{w}_z+\la^2\om_\bot^2w_z=
\frac{3Na_s\hbar^2}{m^2}\frac1{w_\bot^2w_z^2}.
\end{equation}
These equations differ from those in \cite{Perez96} by numerical
factors due to use of the TF distribution in the axial direction
and by absence of the ``quantum pressure'' term in Eq.~(\ref{wz-eq})
which inclusion would be illegal in our approximation. In
dimensionless variables
\begin{equation}\label{q1D-b}
b_\bot=\frac{w_\bot}{a_\bot},\quad b_z=\frac{w_z}{Z_{0}},
\quad \tau=\om_\bot t
\end{equation}
Eqs.~(\ref{wbot-eq}),(\ref{wz-eq}) take the form
\begin{equation}\label{1D-bbot}
\frac{d^2 b_\bot}{d\tau^2}+b_\bot=\frac1{b_\bot^3}+\frac{2q}5
\frac{1}{b_\bot^3b_z},
\end{equation}
\begin{equation}\label{1D-bz}
\frac{d^2b_z}{d\tau^2}+\la^2b_z=
\frac{\la^2}{b_\bot^2b_z^2},
\end{equation}
where the parameter
\begin{equation}\label{1D-q}
q=\left(\frac{\la Z_0}{a_\bot}\right)^2
\end{equation}
measures the strength of the nonlinear pressure in the radial direction.
If $q\ll1$, then we can neglect the second term in the
right-hand side of (\ref{1D-bbot}) and obtain the stationary solution
$b_{\bot 0}=b_{z0}=1$ which corresponds to 1D approximation of Section II.
It is easy to see that in this case Eq.~(\ref{1D-bz}) reduces, for the
case of free axial expansion, to (\ref{b-zdyn}).

The equilibrium values of $b_\bot$ and $b_z$ are determined by
the equations
\begin{equation}\label{1D-equil-mod}
b_{\bot 0}=b_{z0}^{-3/2}, \quad
\frac1{b_{z0}^6}=1+\frac{2q}5\frac{1}{b_{z0}},
\end{equation}
which differ from similar equations of \cite{Das02-1} only through
the notation. They determine the state of BEC before expansion.

Expansion of BEC after switching off the
axial trap is described by the equations
\begin{equation}\label{1D-eqs-mod}
\frac{d^2 b_\bot}{d\tau^2}+b_\bot=\frac1{b_\bot^3}+
\frac{2q}5\frac{1}{b_\bot^3b_z},\quad
\frac{d^2b_z}{d\tau^2}=
\frac{\la^2}{b_\bot^2b_z^2},
\end{equation}
which can be easily solved numerically with the initial conditions
\begin{equation}\label{1D-init-mod}
b_\bot(0)=b_{\bot 0},\quad \dot{b}_\bot(0)=0,
\quad b_z(0)=b_{z0},\quad \dot{b}_z(0)=0,
\end{equation}
where $b_{\bot 0}$ and $b_{z0}$ are determined by Eqs.~(\ref{1D-equil-mod}).
Dependence of $b_\bot=w_\bot/a_\bot$ and $z_{max}=Z_0b_z(t)=Zb_z(t)/b_{z0}$,
where $Z=Z_0b_{z0}$ is the initial half-length of the condensate, is
shown in Fig.~1 for different values of the number of atoms $N$.
\begin{figure}
\includegraphics{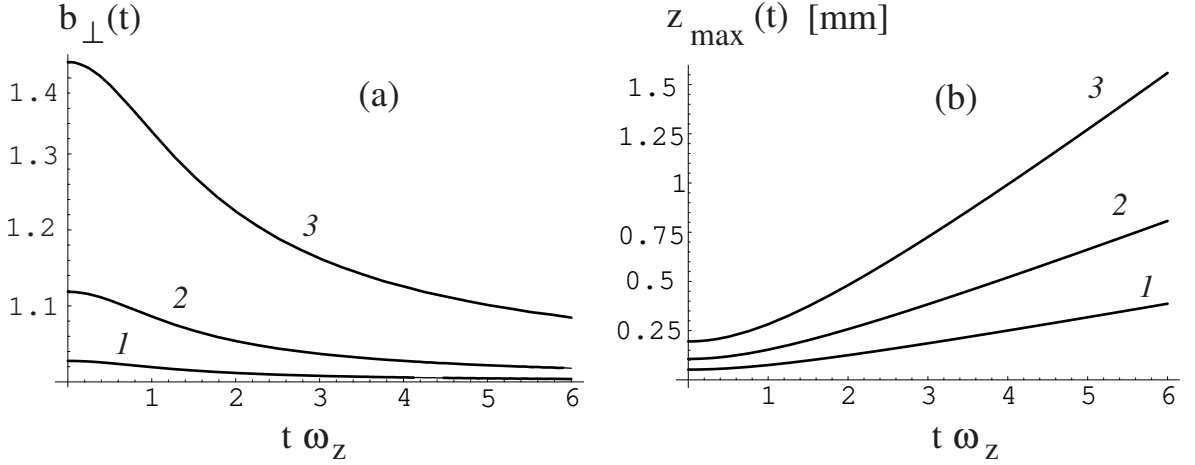}
\caption{Dependence of the dimensionless radial size $b_\bot=w_\bot/a_\bot$ of
the condensate (a) and its axial size $z_{max}$ (b) on dimensionless time
variable
$\tau=t\om_z$ during 1D expansion in a cigar-shaped trap with the
following values of the parameters: $a_\bot=5\,\mu$m, $a_s=5\,$nm,
$\la=0.05$. The curves correspond to the numbers of atoms $N= 10^3$
(1), $10^4$ (2), $10^5$ (3). }
\end{figure}
For $\sqrt{\la}\,a_\bot/a_s\ll N\ll a_\bot/(\la a_s)$ we return to the
analytical solution of Section 2. For $N> a_\bot/(\la a_s)$ evolution
starts from radial width greater than the quantum oscillator length
$a_\bot$ and $b_\bot\to 1$ for $\tau\to\infty$. The axial expansion
for $\tau\gg1$ tends to the motion by inertia with constant velocities
of atoms. It is easy to find the value of maximal velocity with
the use of conservation law of Eqs.~(\ref{1D-eqs-mod}):
\begin{equation}\label{1D-cons}
\frac12\left(\frac{db_z}{d\tau}\right)^2+\frac{5\la^2}{2q}\left[
\left(\frac{db_\bot}{d\tau}\right)^2+b_\bot^2+\frac1{b_\bot^2}\right]
+\frac{\la^2}{b_\bot^2b_z}=\mathrm{const}.
\end{equation}
For $\tau=0$ we have the initial values (\ref{1D-init-mod}) and for
$\tau\to\infty$ we have $db_z/d\tau\to\dot{b}_{z,max}$,
$b_\bot\to1,$ $b_z\to\infty$, hence
\begin{equation}\label{1D-asymp-vel}
\dot{b}_{z,max}=\la\sqrt{
\frac2{b_{\bot 0}^2b_{z0}}+\frac5{q}\left(b_{\bot 0}
-\frac1{b_{\bot 0}}\right)^2}.
\end{equation}
Then with the help of Eqs.~(\ref{1D-equil-mod}) we find
\begin{equation}\label{1D-vmax}
v_{max}=\left.\frac{dz_{max}}{dt}\right|_{t\to\infty}=
\frac{2Z\om_z b_{\bot 0}}{\sqrt{1+b_{\bot 0}^2}}.
\end{equation}
For $q\ll1$, when $b_{\bot 0}=1$, $Z=Z_0$, we return to the 1D formula
(\ref{vmax-as}), and for $q\gg1$, when $b_{\bot 0}\gg1$, we obtain
$v_{max}\simeq 2\om_zZ$. Thus, with increase of the number of atoms
not only longitudinal size $Z$ increases, but also the coefficient
of proportionality between $v_{max}$ and $\om_zR$ increases from
$\sqrt{2}$ to $2$.

Asymptotic distribution of velocities as well as the expression for
the release energy hold their form (\ref{1D-vel-asymp}),(\ref{1D-release}),
where $v_{max}$ is given
by (\ref{1D-vmax}). Dependence of the release energy on the number
of atoms $N$ is shown in Fig.~2.
\begin{figure}
\includegraphics{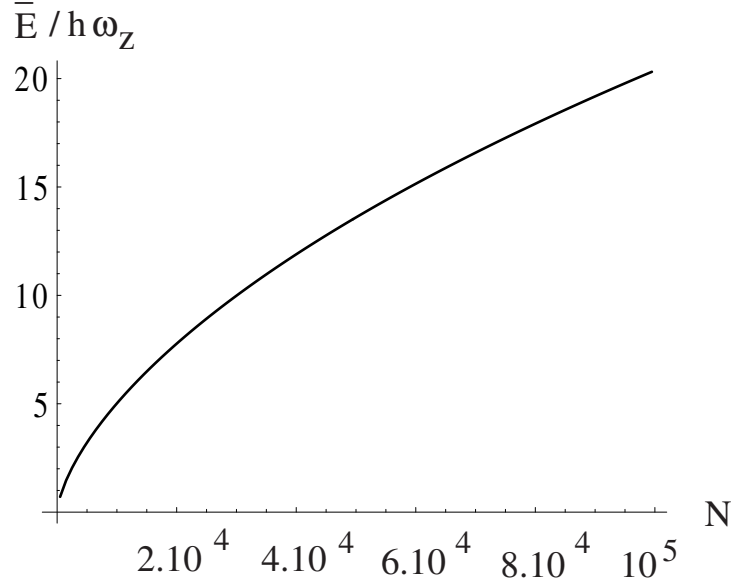}
\caption{The release energy in 1D expansion of the condensate in a
cigar-shaped trap as a function of the number of atoms $N$.
The energy is measured in units of the axial trap oscillator quantum
$\hbar\om_z$.}
\end{figure}

\subsection{Pancake trap}

For pancake trap with TF distribution of density in the transverse
direction, we take the variational trial function in the form
\begin{equation}\label{2D-trial}
\psi=A\sqrt{1-\frac{r^2}{w_\bot^2}}
\exp\left(-\frac{z^2}{2w_z^2}\right)
\exp\left[\frac{i}2(\al_\bot^2r^2+\al_z^2z^2)\right],
\end{equation}
where the time-dependent parameters
$A,$ $w_\bot,$ $w_z,$ $\al_\bot,$ $\al_z$ are connected due to the
normalization condition (\ref{norm3}) by the relation
\begin{equation}\label{2D-A}
A=\left(\frac{2N}{\pi^{3/2}}\frac1{ w_\bot^2w_z}\right)^{1/2}.
\end{equation}
Substitution of (\ref{2D-trial}) into (\ref{action}),(\ref{3Lagr-density})
yields after integration the averaged Lagrangian
\begin{equation}\label{2D-Lagr-av}
\begin{split}
\frac{L}{N}=\left(\frac{\hbar^2}2\frac{d\al_\bot}{dt}+\frac{\hbar^2}{2m}
\al_\bot^2+\frac12m\om_\bot^2\right)\frac{w_\bot^2}3\\
+ \left[\frac{\hbar^2}2\frac{d\al_z}{dt}+\frac{\hbar^2}{2m}\left(
\frac1{w_z^4}+
\al_z^2\right)+\frac12m\om_\bot^2\la^2\right]\frac{w_z^2}2
+\frac8{3\sqrt{2\pi}}
\frac{Na_s\hbar^2}{m}\frac1{w_\bot^2w_z},
\end{split}
\end{equation}
where we have neglected the term $1/w_\bot^4$ which is much less
than $\al_\bot^2$
due to the condition $Na_s/a_z\gg1$ of applicability of the TF approximation
in the radial direction.

The Lagrangian (\ref{2D-Lagr-av}) leads to the relations (\ref{al})
and to equations of motion for widths:
\begin{equation}\label{2D-wbot-eq}
\ddot{w}_\bot+\om_\bot^2w_\bot=\frac{16}{\sqrt{2\pi}}
\frac{Na_s\hbar^2}{m^2}\frac1{w_\bot^3w_z},
\end{equation}
\begin{equation}\label{2D-wz-eq}
\ddot{w}_z+\la^2\om_\bot^2w_z=\frac{\hbar^2}{m^2w_z^3}+
\frac{16}{\sqrt{2\pi}}\frac{Na_s\hbar^2}{m^2}\frac1{w_\bot^2w_z^2},
\end{equation}
which in
dimensionless variables
\begin{equation}\label{q2D-b}
b_\bot=\frac{w_\bot}{R_0},\quad b_z=\frac{w_z}{a_z},
\quad \tau=\om_\bot t,
\end{equation}
where $a_z=a_\bot/\sqrt{\la}$, take the form
\begin{equation}\label{2D-bbot}
\frac{d^2 b_\bot}{d\tau^2}+b_\bot=
\frac{1}{b_\bot^3b_z},
\end{equation}
\begin{equation}\label{2D-bz}
\frac{d^2b_z}{d\tau^2}+\la^2b_z=\frac{\la^2}{b_z^3}+
\frac{\la^2q}3\frac{1}{b_\bot^2b_z^2}.
\end{equation}
The parameter
\begin{equation}\label{2D-q}
q=\left(\frac{R_0}{\la a_z}\right)^2
\end{equation}
measures the strength of the nonlinear pressure in the axial direction.
If $q\ll1$, then we can neglect the second term in the
right-hand side of (\ref{2D-bz}) and obtain the stationary solution
$b_\bot=b_z=1$ which corresponds to 2D approximation of Section II.
It is easy to see that in this case Eq.~(\ref{2D-bbot}) reduces for the
case of free axial expansion to (\ref{2D-eq}).

The equilibrium values of $b_\bot$ and $b_z$ are determined by
the equations
\begin{equation}\label{2D-equil-mod}
b_{z 0}=b_{\bot 0}^{-4}, \quad
\frac1{b_{\bot 0}^{16}}=1+\frac{q}3\frac{1}{b_{\bot 0}^6},
\end{equation}
which again coincide with equations of \cite{Das02-1} up to
the notation. They determine the state of BEC before expansion.
Expansion of BEC after switching off the
radial trap is described by the equations
\begin{equation}\label{2D-eqs-mod}
\frac{d^2 b_\bot}{d\tau^2}=\frac1{b_\bot^3b_z},
\quad
\frac{d^2b_z}{d\tau^2}+\la^2b_z=\frac{\la^2}{b_z^3}+
\frac{\la^2q}3\frac1{b_\bot^2b_z^2},
\end{equation}
which can be solved numerically with the initial conditions
(\ref{1D-init-mod}),
where $b_{\bot 0}$ and $b_{z0}$ are determined by eqs.~(\ref{2D-equil-mod}).
The dependence of $R_{max}=Rb_\bot(t)/b_{\bot 0}$ and $b_z=w_z/a_z$,
where $R=R_0b_{\bot 0}$ is the initial radius of the condensate, is
shown in Fig.~3 for different values of the  number of atoms $N$.
\begin{figure}
\includegraphics{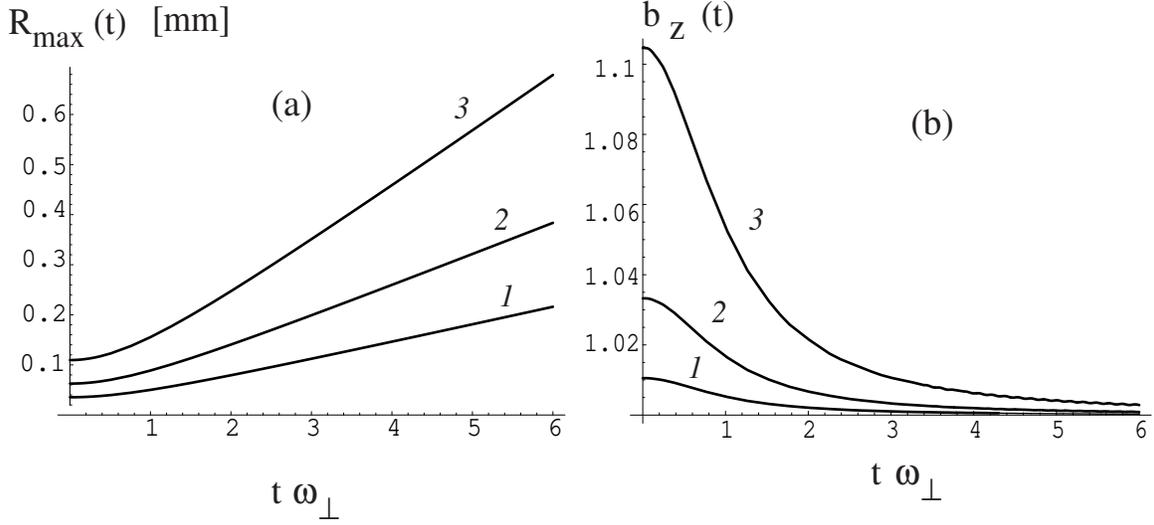}
\caption{Dependence of the radial size $R_{max}$ of
the condensate (a) and its dimensionless axial size
$b_z= w_z/a_z$ (b) on dimensionless time variable
$\tau=t\om_\bot$ during 2D expansion in a pancake trap with the
following values of the parameters: $a_z=5\,\mu$m, $a_s=5\,$nm,
$\la=20$. The curves correspond to the numbers of atoms $N= 10^3$
(1), $10^4$ (2), $10^5$ (3). }
\end{figure}
For $a_z/a_s\ll N\ll a_z\la^2/a_s$ we return to the
analytical solution of Section 2. For $N> a_z\la^2/a_s$ the evolution
starts from axial width $w_{z0}$ greater than the quantum oscillator length
$a_z$ (i.e. $b_{z0}>1$) and $b_z\to 1$ for $\tau\to\infty$. The radial
expansion
for $\tau\gg1$ tends to the motion by inertia with constant velocities
of atoms. The maximal radial velocity can be found with
the use of conservation law of Eqs.~(\ref{2D-eqs-mod}),
\begin{equation}\label{2D-cons}
\left(\frac{db_\bot}{d\tau}\right)^2+\frac{3}{\la^2q}\left[\frac12
\left(\frac{db_z}{d\tau}\right)^2+\frac{\la^2}2\left(b_z^2+\frac1{b_z^2}
\right)\right]
+\frac{1}{b_\bot^2b_z}=\mathrm{const}.
\end{equation}
which yields
\begin{equation}\label{2D-asymp-vel}
\dot{b}_{\bot,max}=\sqrt{
\frac1{b_{\bot 0}^2b_{z0}}+\frac3{2q}\left(b_{z 0}
-\frac1{b_{z 0}}\right)^2}.
\end{equation}
Then for maximal velocity we obtain
\begin{equation}\label{2D-vmax}
v_{max}=\left.\frac{dR_{max}}{dt}\right|_{t\to\infty}={\om_\bot R}
\sqrt{\frac{1+3b_{z 0}^2}{2(1+b_{z 0}^2)}}.
\end{equation}
For $q\ll1$, when $b_{z0}=1$, $R=R_0$, we return to the 2D formula
(\ref{asymp}), and for $q\gg1$, when $b_{z 0}\gg1$, we obtain
$v_{max}\simeq \sqrt{3/2}\,\om_\bot R$.
Asymptotic distribution of velocities as well as the expression for
the release energy hold their form (\ref{a}),(\ref{b}),
where $v_{max}$ is given
by (\ref{2D-vmax}). Dependence of the release energy on the number
of atoms $N$ is shown in Fig.~4.
\begin{figure}
\includegraphics{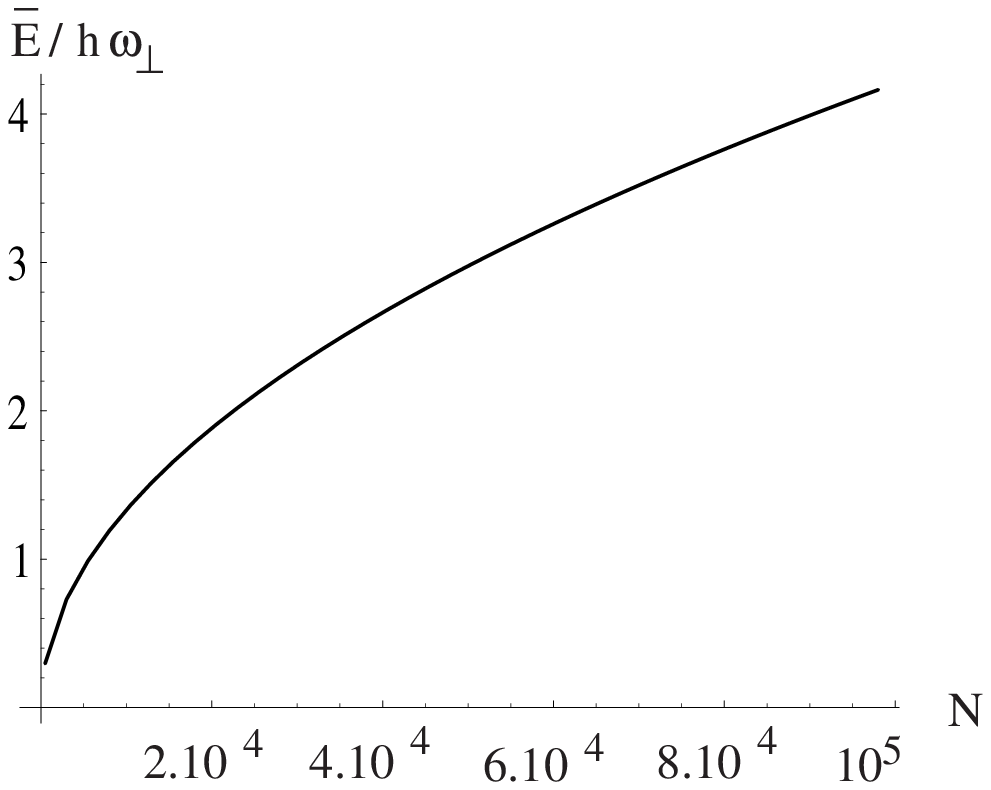}
\caption{The release energy in 2D expansion of the condensate in a
pancake trap as a function of the number of atoms $N$.
The energy is measured in units of the radial trap oscillator quantum
$\hbar\om_\bot$.}
\end{figure}

\section{Discussion}

Here we compare the theory with available experimental data.

Expansion of the condensate in quasi-1D waveguide was studied
experimentally in \cite{Bongs01}. In this experiment, the half-length
of the condensate axial size was $Z\simeq 100\,\mu$m, and the time of
transition to expansion by inertia was about $20\,$ms which gives
$\om_z\simeq 50\,$s$^{-1}$. Then formula (\ref{vmax-as}) gives the value
of the maximal velocity
$$
v_{max}\simeq \sqrt{2}Z\om_z\simeq 7\,\mathrm{mm/s}
$$
which, in view of our approximate estimates of relevant parameters,
agrees very well with the experimental value
$v_{max}\simeq 5.9\,$mm/s.

 Free expansion of BEC in an array of 2D
traps has been investigated in \cite{Morsch02}. It is clearly seen in Fig.~2 of
this paper that no axial expansion is observed what means that
the condensate is effectively split into several independent
condensates confined in the individual lattice wells.
Therefore we can apply 2D theory to description of free
radial expansion of each separate condensate. From Fig.~2
of \cite{Morsch02} we find that maximal radial velocity
is equal to $v_{max}\simeq 1.5-1.7\,$mm/s, the initial radius
is $R\simeq 13\,\mu$m, and the radial trap frequency is
$\om_\bot\simeq 132\,\mathrm{s}^{-1}$. Then formula (\ref{asymp})
gives
$$
v_{max}\simeq R\om_\bot\simeq 1.7\,\mathrm{mm/s}
$$
in excellent agreement with the above experimental value.
Thus, as we see, the theory agrees  with experimental data. Experimentally
observed deviations from theoretical predictions could serve
as indications on realization of non-mean-field regimes of BEC
behavior.

In conclusion, we have considered in detail the theory of one-
and two-dimensional expansion of BEC. Three-dimensional effects
have been estimated with the use of variational approach.
Good agreement with the available experimental data is found.

\subsection*{Acknowledgements}

I am grateful to F.Kh.~Abdullaev, V.A.~Brazhnyi,  A.~Gammal,
V.V.~Konotop, R.A.~Kraenkel,
and L. Tomio for valuable discussions.
I thank also the staff of Instituto de F\'{\i}sica
Te\'{o}rica for kind hospitality. This work was
supported by FAPESP (Brazil).

\end{document}